\documentclass[10pt,preprint2]{aastex}
\usepackage{amsmath,amssymb,graphicx,soul,color}
\usepackage{epstopdf}
\usepackage{hyperref}
\usepackage{soul}
\usepackage{float}
\usepackage{booktabs}

\usepackage{natbib}
\bibpunct{(}{)}{;}{a}{,}{,}
\begin{document}

\title{GRB Flares:  UV/Optical flaring (Paper I)}

\author{C. A. Swenson\altaffilmark{1}, P. W. A. Roming\altaffilmark{2,1}, M. De Pasquale\altaffilmark{3}, S. R. Oates\altaffilmark{3}}

\email{cswenson@astro.psu.edu}
\altaffiltext{1} {Pennsylvania State Univ., 525 Davey Lab, University Park, PA 16802, USA}
\altaffiltext{2}{Southwest Research Institute, 6220 Culebra Road, San Antonio, TX 78238, USA}
\altaffiltext{3}{Mullard Space Science Laboratory, University College London, Surrey, UK}

\begin{abstract}

We present a new algorithm for the detection of flares in gamma-ray burst (GRB) light curves and use this algorithm to detect flares in the UV/optical.  The algorithm makes use of the Bayesian Information Criterion (BIC) to analyze the residuals of the fitted light curve, removing all major features, and to determine the statistically best fit to the data by iteratively adding additional `breaks' to the light curve.  These additional breaks represent the individual components of the detected flares: $T_{start}$, $T_{stop}$, and $T_{peak}$.  We present the detection of 119 unique flaring periods detected by applying this algorithm to light curves taken from the Second \emph{Swift} Ultraviolet/Optical Telescope (UVOT) GRB Afterglow Catalog.  We analyzed 201 UVOT GRB light curves and found episodes of flaring in 68 of the light curves.  For those light curves with flares, we find an average number of $\sim$2 flares per GRB.  Flaring is generally restricted to the first 1000 seconds of the afterglow, but can be observed and detected beyond $10^{5}$ seconds.  More than 80\% of the flares detected are short in duration with $\Delta t/t$ of $< 0.5$.  Flares were observed with flux ratios relative to the underlying light curve of between 0.04 to 55.42.  Many of the strongest flares were also seen at greater than 1000 seconds after the burst.

\end{abstract}


\section{Introduction}

One of the many great advances made by the \emph{Swift} \citep{Gehrels2004} mission was that of early time Gamma-ray Burst (GRB) afterglow follow up.  \emph{Swift} was specifically designed with rapid GRB afterglow observations in mind.  Prior to the launch of \emph{Swift}, GRB afterglow observations generally did not start until hours after the burst, and an X-ray position was generally needed before any optical follow-up could occur.  This meant that most optical detections did not take place until days after the GRB.  \emph{Swift} would solve this problem through the use of 3 separate instruments on the same spacecraft working together.  After the detection of a GRB by the Burst Alert Telescope (BAT; \citealt{Barthelmy2005}), \emph{Swift} autonomously slews to the position, generally within $\sim$ 100 seconds, allowing the X-ray Telescope (XRT; \citealt{Burrows2005}) and UV/Optical Telescope (UVOT; \citealt{Roming2000},\citeyear{Roming2004},\citeyear{Roming2005}) to begin observations of the afterglow.  \emph{Swift} has proven to be invaluable in furthering our understanding of GRB physics, having observed over 700 GRBs, and was specifically designed to be able to observe the early afterglow evolution and transition from the prompt emission to the afterglow stage.  The early stages of the afterglow proved to be very exciting and led to the discovery of a number of new features, including the ``canonical" X-ray light curve \citep{Nousek2006}, which has been observed in a number of GRBs (e.g. \citealt{Hill2006}, \citealt{Evans2009}).  Another important feature seen early in the \emph{Swift} mission was X-ray flares (e.g. \citep{Burrows2005a,Romano2006}).

Flares in X-ray light curves had been seen prior to their discovery in XRT light curves (e.g. \citealt{Piro1998,Piro2005}), but had only been observed a handful of times.  It was quickly shown that they are quite common, appearing in approximately 50\% of XRT afterglows \citep{OBrien2006}, and are temporally displaced so as to be distinct from the prompt emission.  Flares are observed as superimposed deviations from the underlying light curve and have been observed in all phases of the canonical X-ray light curve.

Early in the \emph{Swift} mission, several studies were performed that highlighted individual GRBs that exhibited either large numbers of flares or flares of unusually high fluence.  Each of these studies expanded our understanding of flares and the physical processes whereby they are created.  In particular, the studies of GRBs 050406 \citep{Romano2006}, 050502B \citep{Falcone2006}, 050713A \citep{Morris2007}, 050724 \citep{Campana2006} and 050904 \citep{Cusumano2007} established the fact that flares are likely caused by internal shocks because of their steep rise and decay slopes, though the actual source of the flares is still debated and may be linked to instabilities in the ejecta or the release of stored electromagnetic energy.  These studies also showed that X-ray flares are observed in both long and short GRBs, can contain energies as large as the prompt emission, they appear to come from a distinct emission mechanism than the afterglow, are seen in both long and short GRBs, and can be temporally separated from the prompt phase by hundreds of seconds.  Further studies have only reinforced these initial findings and have even shown that significant flares can be created at times greater than $10^5$s after the initial prompt detection (e.g. \citealt{Swenson2010}).

Some attempts have been made to look at larger collections of flares and have examined their properties on a more generalize basis.  \citet{Falcone2007} and \citet{Chincarini2007} examined the temporal and spectral properties, respectively, from a collection of flares found in 33 of the first 110 GRBs observed by \emph{Swift}.  The combined results from these two studies found that the late-time internal shocks were \emph{required} to explain 10 of the observed flares and that central engine activity was the preferred method for a majority of the bursts.  However, \citet{Chincarini2007} also state that more observations of flares over more energy bands are needed.  A follow-up study was performed \citep{Chincarini2009} that limited the data set to to those GRBs which had redshifts, enabling a study of the actual energetics of the flares and found some indication that the flare energy may be correlated to the GRB prompt energy, but was limited due to the number of bursts.  They once again confirmed that more observational work is needed.  Additional studies further confirmed earlier results showing that X-ray flares are likely caused by late-time internal dissipation processes, which produces the prompt emission, and also showed that flares evolve over time, becoming broader and flatter.  However these studies limited their data to only the first 1000 seconds of the GRB afterglow light curve \citep{Chincarini2010} or a limited sample of 9 exceptionally bright X-ray flares \citep{Margutti2010}.

An attempt at incorporating information from multiple energy bands was made by \citet{Morris2008} in which spectral energy distributions (SEDs) were created, using BAT, XRT and UVOT data, for flares found in the same sample of 110 GRBs used by \citet{Falcone2007} and \citet{Chincarini2007}.  The fits to the SEDs showed that the flares, unlike the afterglow, could not be fit by a simple absorbed power law.

The number of studies analyzing flares in the UV/optical are even more limited than those for the X-ray \citep{Roming2006a}. The primary reason for this is the lower significance of most flares in the lower energy bands.  While the X-ray flares are often easily identified by visual inspection of the light curves, potential UV/optical flares are more often overlooked or dismissed as noise.  

A notable example of flares detected by the UVOT is found in the light curve of the short GRB 060313 \citep{Roming2006b} in which late-time flaring was observed by the UVOT, but not seen in the XRT.  The type of behavior seen in GRB 060313 are not unusual to the study of flares in GRBs.  The presence of flares in one bandpass, but not in another is seen often, and just as common are cases where the flares are observed from X-ray all the way to the optical.  In the specific case of GRB 060313 the flares could be consistent with density fluctuation in the circumstellar medium, provided that the cooling frequency, $\nu_c$, lies between the X-ray and UV/optical bands, which would explain why the flares appeared in the UV/optical but not in the X-ray.  However, the flares can also be explained by central engine activity at late times, which is the more preferred explanation for X-ray flares as stated above.  Another notable example is that of GRB 090926A \citep{Swenson2010}.  This burst displayed the previously mentioned late time flares at times greater than $10^5$ seconds, which can be explained by central engine activity at \emph{extremely} late times, but also because the flaring is simultaneously observed in the X-ray as well as the UV/optical.  Identifying the source of the flares and whether X-ray and UV/optical flares have the same origin remains an important open question.

The common factor in all of the aforementioned studies is that the flares were found by simple manual inspection of the light curves and were easily detectable by eye.  This method has allowed for a significant number of X-ray flares to be detected, but has yielded a very small number of UV/optical flares due to the difficulties listed above.  A blind, systematic search for flares in both X-ray and UV/optical bands has not yet been performed and is necessary to provide an unbiased sample of flares of all brightnesses.  Such a sample would be able to address some of the limitations mentioned in the previous X-ray studies and would provide access to a relatively untapped source of knowledge with additional UV/optical flares.  

The complementary nature of two such flare catalogs would allow for more stringent constraints on the origin of flares in GRBs through cross-correlation of the two energy regimes.  The precise nature of the GRB central engine still remains largely unknown and, because flaring is most likely related to central engine activity, the study of flares is crucial to our unlocking of that mystery.

In this paper we present the results from a blind, systematic search for flares in UVOT light curves.  Using Monte Carlo simulations and dynamic programming in conjunction with Bayesian statistics we have constructed the most complete catalog of UV/optical flares to date, and provide the temporal details of each flare, including $T_{peak}$, $\Delta t/t$, and the strength of the flare relative to the underlying light curve.  In a forthcoming paper we will perform a similar analysis on the matching XRT dataset and present our joint analysis of the two catalogs.

This paper is organized as follows:  In \S 2 we describe our data set as well as our methodology for identifying flares.  We present our catalog of GRB flares observed by the UVOT in \S 3, and discuss the implications drawn from the catalog in \S4.  In \S5 we present our future plans for utilizing this new flare detection algorithm.

\section{Methodology}

For the purposes of this study we will be using the light curves produced by \citet{Roming2013}.  This Second \emph{Swift} Ultraviolet/Optical Telescope GRB Afterglow Catalog provides a complete dataset of fitted UVOT light curves for both long and short  GRBs observed by \emph{Swift} from April 2005 through Dec 2010, and makes use of optimal co-addition \citep{Morgan2008} which results in better sampled light curves and increases the probability of detecting flares.  In addition to being optimally co-added, the Second UVOT GRB Catalog also normalizes the GRB light curves to a single filter from the 7 possible filters used during observations.  This normalization further increases the timing resolution of the overall light curve during those periods when multiple filters were used during the same orbit.  These two methods combined, optimal co-addition and normalization, have resulted in a completely unique and previously unavailable data set that is suited for use in searching for flares and other small features \citep[as opposed to the First UVOT GRB Catalog; ][]{Roming2009}.  This sample is also approximately twice as large as the sample provided in \citet{Roming2009}.  Due to the normalization of the light curves, we will not be performing any chromatic analysis on the light curves, and our experience in fitting GRB light curves leads us to believe that our detected flares will not evolve significantly with energy over the limited energy band observed by the UVOT.  Our analysis described below is performed on the residuals of the fitted UVOT light curves, using the fitting parameters provided by \citet{Roming2013} when calculating our residuals.

Even with the increased probability of detecting flares that comes with using optimally co-added data, the flares that we hope to identify are below the significance level of the previous X-ray studies previously mentioned (e.g. \citealt{Falcone2007,Chincarini2007}) and we require a statistically robust method to confirm that the flares are real and not part of the background noise.  For this study we have used the publicly available \texttt{R} package \texttt{strucchange} \citep{Zeileis2002} and the breakpoints function contained within the package \citep{Zeileis2003}.  The breakpoints function specifically employs the approach of dynamic programming to compute the optimal number of breakpoints in a time series of data.  In terms of GRB light curves, a breakpoint is defined as the time at which the subsequent data are no longer fit by the preceding slope (i.e. the beginning or ending of a flare, or any other break in the light curve).  The computation of breakpoints involves breaking the time series into smaller pieces, iteratively fitting the curve and minimizing the residual sum of squares (RSS).  As one would expect, simply minimizing the RSS would lead to fitting the light curve with $n - 1$ breakpoints (where $n$ is the number of data points), with an individual segment connecting each data point.  To counter this effect, the breakpoints function also computes the Bayesian information criterion \citep[BIC; ][]{Schwarz1978},

\begin{equation}
BIC = -2 \times L + k\ln(n),
\end{equation}

where $k$ is the number of free parameters to be estimated, $n$ is again the number of data points in the light curve and $L$ is the maximized value of the log-likelihood function,

\begin{equation}
L = (\log p(D|\theta_{j}, M_{j}) - \log p(D|\theta_{j+1},M_{j+1})),
\end{equation}

which compares the probabilities of two possible fits to the data, functions $M_{j}$ and $M_{j+1}$, each with their respective set of parameters $\theta_{j}$ and $\theta_{j+1}$, and returns the more likely fit based on the observed data points, $D$.  The BIC is penalized and becomes increasingly large when either the data is poorly fit or when the number of free parameters increases and the data is overfit.  The BIC is therefore minimized using the simplest model that sufficiently fits the data.  The breakpoints function returns the \emph{optimal} number of breakpoints by calculating the fit that minimizes the combined RSS and BIC values.

The BIC is unlike many of the more commonly used statistical measurements used in astronomy (e.g. $\chi^{2}$, F-test, etc.).  The BIC is a ``summary of the evidence provided by the data in favor of one scientific theory, represented by a statistical model, as opposed to another" \citep{Kass1995}, but it does not provide a definite strength or probability to a preferred statistical model.  \citet{Kass1995} do provide a guideline for interpreting the strength of evidence for the preferred fit:

\begin{center}
    \begin{tabular}{lp{1.6in}}
        $BIC_{i}$ $-$ $BIC_{min}$ & Evidence Against Model $i$      		\\ 
	\hline
        0 to 2				 & Not worth more than a bare mention     \\ 
        2 to 6				 & Positive          			                \\ 
        6 to 10				 & Strong                            			 \\ 
        $>$10				 & Very Strong                      			 \\
      \end{tabular}
\end{center}

For our purposes we will require $BIC_{i} - BIC_{min} > 6$, or `Strong' evidence, to determine the preferred fit.  In the case of $BIC_{i} - BIC_{min} < 6$, we will adopt the simplest fit model (i.e. fewest breakpoints) that satisfies the criterion of $BIC_{i-1} - BIC_{i} > 6$.

The breakpoints function does not take into account the systematic and random error present in all data.  In order to reintroduce the effects of these errors, we run a Monte Carlo simulation, randomizing the values of the observed data points in line with the measured errors.  For each Monte Carlo iteration we calculate the optimal number of breakpoints and their corresponding times.  For the purposes of this study we performed $10,000$ Monte Carlo simulations on each GRB light curve.  For each light curve we examine the BIC value for each potential number of breakpoints (i.e. $1, 2, 3, 4 ... n-1$) over all $10,000$ iterations and do the same for the RSS values.  Applying the $BIC_{i} - BIC_{min} > 6$ criteria allows us to determine the statistically preferred fit to the data, that does not overfit the light curve.  Each of the  breakpoints found in this optimal fit correspond to a specific data point where the original fit to the GRB light curve no longer adequately describes the data and is likely caused by a flare.  Appendix A provides a more detailed explanation, showing results of the breakpoints function running on simulated light curves and flares.

\begin{figure}
	\includegraphics[width=0.8\textwidth,angle=90]{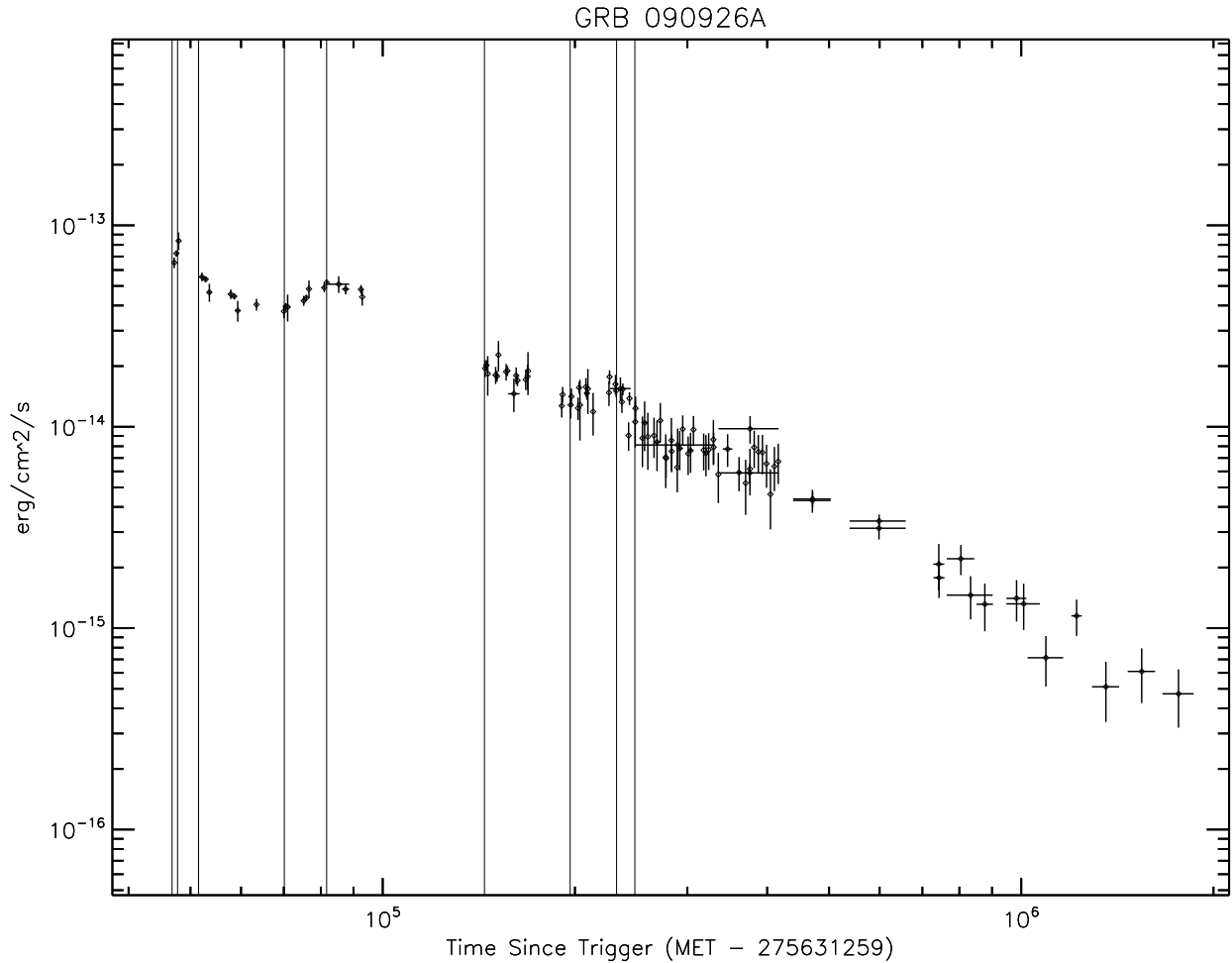}
	\caption{UVOT light curve of GRB 090926A showing the 9 breakpoints detected by the flare finding algorithm.  The 9 breakpoints are combined to form 3 individual flares.}
	\label{fig:GRB090926A}
\end{figure}

Once these potential flares have been identified, we extract the properties of the flare from the light curve.  A well defined flare consists of three breakpoints:  1)  $T_{start}$, the time of the initial deviation from the underlying light curve decay, 2) $T_{peak}$, the time of the peak of the flare, where the slope goes from positive to negative again, and 3) $T_{stop}$, the time when the decay of the flare returns to the underlying decay slope.  In most cases we can not precisely identify the exact time of $T_{start}$, $T_{peak}$, or $T_{stop}$ due to insufficient sampling of the light curve, or observing constraints creating gaps in the light curve.  Because of these limitations we only calculate limits on the boundaries of $T_{start}$ and $T_{stop}$ based on the available data, and cannot precisely define $T_{peak}$, but rather use the observed data point exhibiting the highest flux during the period of flaring as a lower limit.  Our estimates of the peak flux ratio, relative to the underlying light curve, will also be a lower limit due to the limitations in calculating $T_{peak}$.  This approach will ensure that we do not bias further studies with underestimated values of the peak flux.  Our determination of $\Delta t/t$ is also limited by the uncertainty in determining $T_{peak}$ because the true peak flux may occur any time between $T_{start}$ and $T_{stop}$, but because most flares are relatively short the error in our estimate of $\Delta t/t$ is only a few percent.   In the few cases of flares with distinct features that could be analytically fit (e.g. $T_{start}$ and $T_{peak}$ for the flare peaking near 80 kiloseconds in GRB 090926A shown in Figure~\ref{fig:GRB090926A}), these limits closely match the values derived from fitting the flare itself.  We therefore see no need to apply a different approach to the those few exceptional flares, by explicitly fitting them, but rather use the same limit approach as for the rest of the data set.

Figure~\ref{fig:GRB090926A} shows the results of running the flare finding algorithm on the light curve of GRB 090926A.  Because the flaring occurs at such late times, the flares are stretched out and the individual components become easy to see.  The figure show the 9 identified breakpoints as vertical lines.  These 9 breakpoints were then combined to create 3 individual flares, each with a $T_{start}$, $T_{peak}$ and $T_{stop}$.  The observing gap after the peak of the second flare means that we can only provide a lower limit on $T_{stop}$ by placing it at the first data point after the gap.  Using the flare finding algorithm we were able to successfully identify the two flares that we previously identified \citep{Swenson2010}, but also detected a third flare starting at the beginning of the light curve that we were unsure of when attempting to identify flares by eye.

\section{Results}

Here we present the results of our analysis of the 201 UVOT GRB light curves from the Second UVOT GRB Catalog.  We find the presence of at least 119 unique flares, detected in 68 different light curves.  Table~\ref{tab:Flaretable}) provides the following information for each flare:  1) GRB Name, 2) the flare peak time, defined as the observed time since the initial burst of the highest flux data point during the flaring period, as well as the limits on 3) $T_{start}$ and 4) $T_{stop}$, set to the last and first data points, respectively, that are well fit by the underlying light curve.  5) A limit on $\Delta t/t$ based on the calculated peak time, $T_{start}$ and $T_{stop}$, and 6) a lower limit on the ratio of the peak flux during the flaring period, relative to the underlying light curve, using the observed flux at the flare peak time.  The flux ratio is normalized using the flux of the underlying light curve to allow for direct comparison of each flare across all light curves.  Lastly, 7) the confidence measure of the detected flare as a fraction of the number of times the flare was identified during the $10,000$ Monte Carlo simulations.

\section{Discussion}

Our analysis shows that at least 33\% of the light curves in the Second UVOT GRB Catalog contain possible episodes of flaring.  This number is very much in line with analysis that has been performed on X-ray light curves (e.g. \citealt{Chincarini2010}).  This result, however, does not correct for those light curves that were so poorly sampled, or for which observations did not start until such late times as to make the detection of any flares challenging.  It is not unreasonable to assume that an even larger fraction of the light curves in the Second UVOT GRB Catalog contain flares that will simply never be detected due to these issues.

For the purposes of this analysis we have divided the detected flaring period into three groups.  First the `gold' group, defined as those flares with a confidence measure greater than 0.7 and a $\Delta t/t \le 0.5$.  This group provides a good detection rate and satisfies the somewhat `classic' definition of a flare.  This group contains 46 flares.  Next is the `silver' group, which expands the parameters to flares with confidence greater than 0.6 and $\Delta t/t \le 1.0$.  This group includes the 46 `gold' flares and adds an additional 24, for a total of 70 flares.  The final group, the `bronze' group, contains all 119 detected flares.

\begin{figure}
	\includegraphics[scale=0.6,angle=-90]{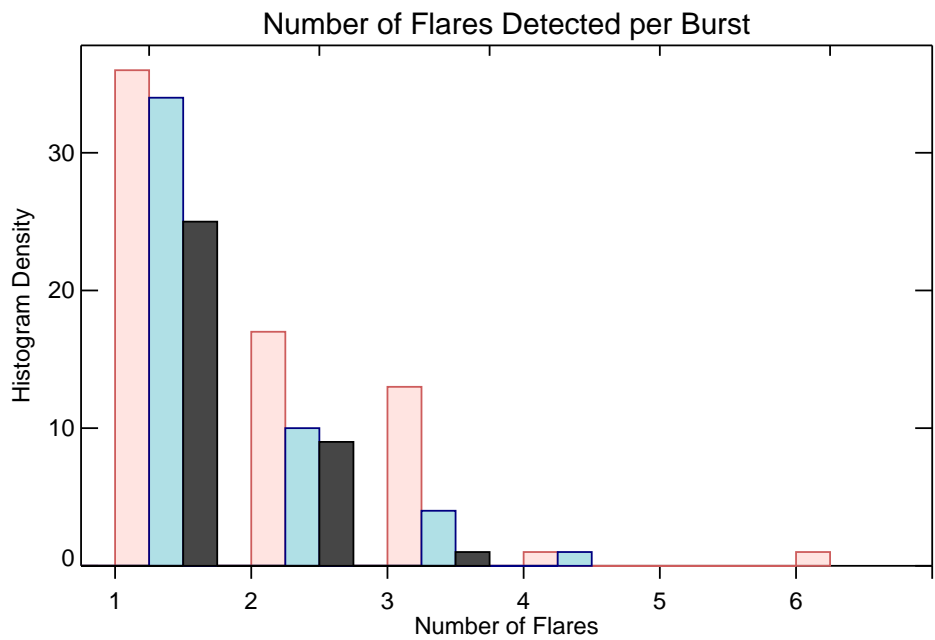}
	\caption{Distribution of the number of detected flares per GRB.  The three distributions are the gold (black font), silver (blue font) and bronze (red font) distributions described in the text.}
	\label{fig:Number_of_Flares_Histogram}
\end{figure}

For those light curves with detected flares, the most common number of flares for each group is 1 per GRB with the average number of flares for each group being $\sim$2.  Fig~\ref{fig:Number_of_Flares_Histogram} shows the distribution of flares per GRB for the gold, silver, and bronze groups, shown in black, blue and red, respectively.  The flares in the gold and silver groups come primarily GRBs with a single flare, while the bronze group has a large fraction of its flares coming from GRBs with two or three flares.  No GRB had more than three flares all belonging to the gold group.  GRB 090618 displayed the most flaring activity with 6 unique flaring episodes detected, with all of its flares belonging to the bronze group.

\begin{figure}
	\includegraphics[width=0.6\textwidth,angle=-90]{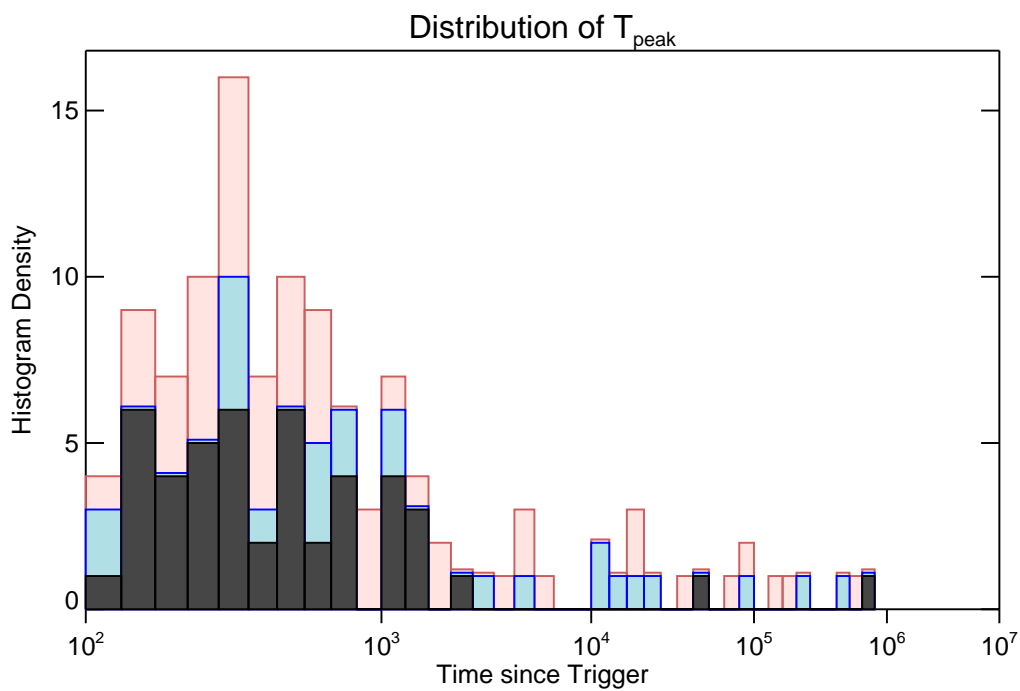}
	\caption{Histogram of the distribution of $T_{peak}$.  The three distributions are the gold (black font), silver (blue font) and bronze (red font) distributions described in the text.  Note:  In cases where no additional flares added when going from the gold to silver group (or from silver to bronze) the bin of the former group is shown as a thin line on top of the latter group.}
	\label{fig:T_peakHistogram}
\end{figure}

The earliest flare peak time occurs at 108 seconds after the trigger of GRB 060708, and the latest detected flare peaks at 787 kiloseconds after the trigger of GRB 050712, with 85\% of all detected flares peaking before 1000 seconds.  Figure~\ref{fig:T_peakHistogram} shows the distribution of $T_{peak}$ for the three groups of flares.  The gold group has an average $T_{peak}$ of $\sim$500 seconds, while the introduction of later flares in the silver and bronze groups push their average $T_{peak}$ to $\sim$900 seconds after the GRB trigger.  However, the most commonly observed $T_{peak}$ for both the silver and bronze groups is also $\sim$500 seconds.  It appears that all three groups are similarly distributed, but with fewer late time detections in the gold group due to the strict criteria for gold designation and the difficulties in detecting flares at late times.  The distributions for all three groups can be reasonably fit by the same Gaussian, with different maximums, centered around $\sim$300 seconds with a full width at half max of 10 seconds.

\begin{figure}
	\includegraphics[width=0.6\textwidth,angle=-90]{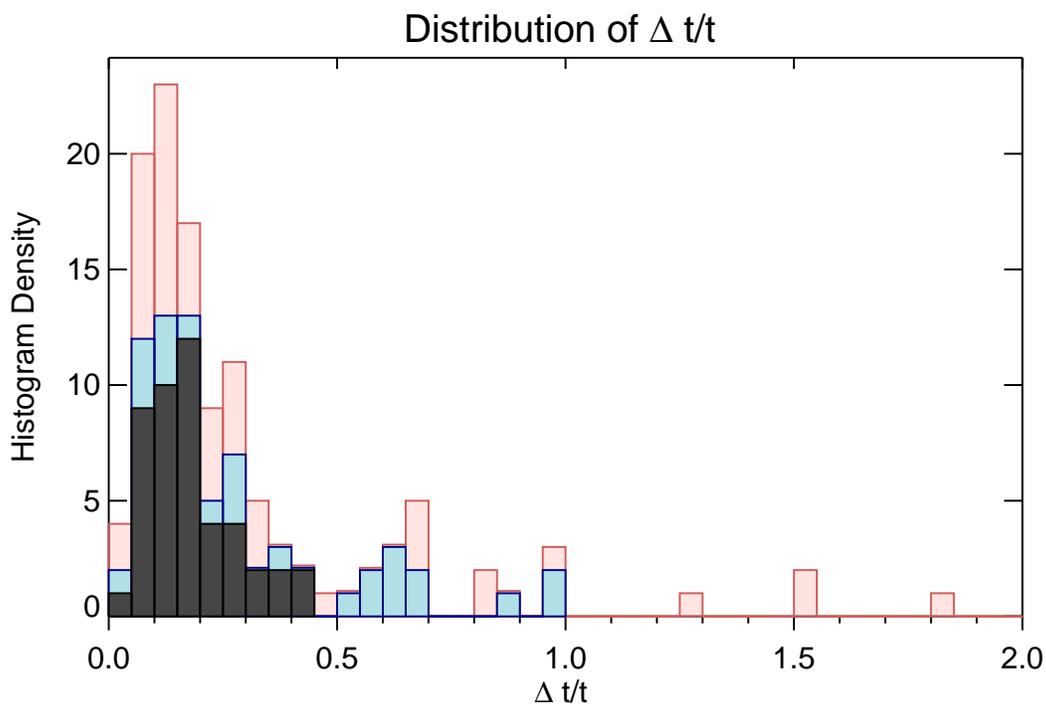}
	\caption{Distribution of $\Delta t/t$ for the detected flares.  The three flares with $\Delta t/t > 2.0$ are omitted for scaling purposes.  The three distributions are the gold (black font), silver (blue font) and bronze (red font) distributions described in the text.  Note:  In cases where no additional flares added when going from the gold to silver group (or from silver to bronze) the bin of the former group is shown as a thin line on top of the latter group.}
	\label{fig:delta_t_t_Histogram}
\end{figure}

The duration of the flares, taking into account the limited nature of our determination of $T_{start}$ and $T_{stop}$, vary from $\Delta t/t$ of 0.01 to greater than 10, with at least 80\% of the bursts exhibiting a $\Delta t/t < 0.5$.  Figure~\ref{fig:delta_t_t_Histogram} shows the distribution of $\Delta t/t$ for the three groups of bursts.  As with $T_{peak}$, all three groups can be fit by the same Gaussian function, with different maximums, centered at 0.14 seconds and with a full width at half max varying between 0.17 seconds for the gold group to 0.21 seconds for the bronze group.  The consistency between the three groups for both $T_{peak}$ and $\Delta t/t$ shows that the detection algorithm is robust even at lower confidence levels.  It should be noted that for flares with $\Delta t/t > 0.5$, particularly those observed in the first few hundred seconds of the light curve, may actually be the onset of the forward shock emission \citep{Oates2009}.

\begin{figure}
	\includegraphics[width=0.6\textwidth,angle=-90]{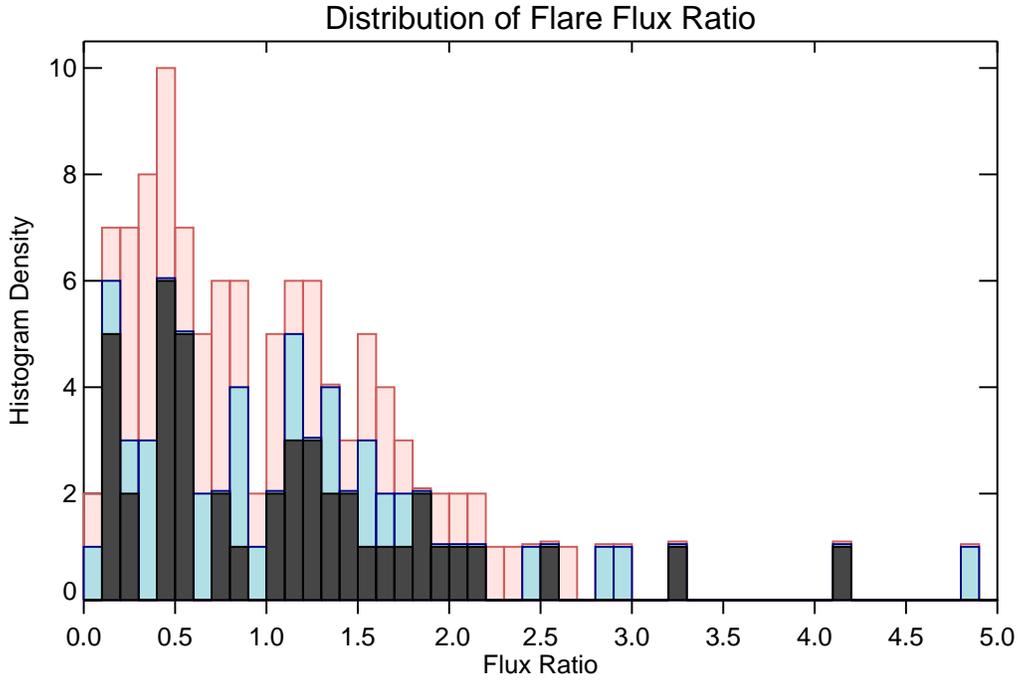}
	\caption{Distribution of flare flux ratio, relative to the underlying light curve.  The flares with flux ratios $> 5$ are omitted for scaling purposes.  The three distributions are the gold (black font), silver (blue font) and bronze (red font) distributions described in the text.  Note:  In cases where no additional flares added when going from the gold to silver group (or from silver to bronze) the bin of the former group is shown as a thin line on top of the latter group.}
	\label{fig:Flux_Ratio_Histogram}
\end{figure}

The relative strengths of the flares varies from a minimum flux ratio of $0.04$ to a maximum of $55.42$.  Figure~\ref{fig:Flux_Ratio_Histogram} shows the distribution of the flare flux ratios for the three groups.  The bronze and silver groups can be fit by a Gaussian centered around $\sim$0.5, whereas the gold would be better fit by a series of Gaussians due to the valleys between the peaks at $\sim$0.2, $\sim$0.5 and $\sim$1.3.  This may indicate a preferred set of flare strengths in the UV/optical, though, a single Gaussian fit to the gold group does provide a centered value consistent with the bronze and silver groups, indicating that it is likely a lack of data causing the poor fit and that all three groups are in fact equally well fit by the same Gaussian.  A larger sample of gold flares, meaning continued GRB light curve observations in the UV/optical, will be required to determine whether it is a lack of data or an actual physical phenomenon causing the observed structure.  More than 83\% of the bursts have flux ratios between $0.04$ and $2$, with representatives in each of the three groups, while there are 19 relatively strong flares with flux ratios $> 2$.  Interestingly, 14 of the 19 relatively strong flares are among the ~15\% of flares that peak later than 1000 seconds.  After the first 1000 seconds, light curves have generally poor timing resolution due to the decaying nature of the afterglow, so it may be a simple observational bias that leads to a large fraction of those late-time flares having large flux ratios (i.e. larger flares are easier to to detect than small flares at late times).  Further analysis and simulations will be required to determine whether an observational bias exists or whether the high fraction of large late-time flares is indicative of a unique subset of GRBs capable of producing these types of flares.

In addition to analyzing the burst parameters individually, we also performed an analysis of the UV/optical flare parameters compared to the GRB prompt parameters for each burst.  Specifically we compared the reported $T_{90}$, prompt emission fluence, and the amount of structure present in the prompt emission (i.e. single FRED-like peak versus multi-peak structure) to $T_{peak}$, $\Delta t/t$, the flux ratio and the number of flares per GRB and find no correlation between any of the prompt emission parameters and flare parameters.  We interpret the lack of correlation to indicate that the emission source of the UV/optical flares detected is not the same as that of the high energy prompt GRB emission.

\section{Future Work}

We have introduced a new flare detection algorithm and have analyzed the GRB light curves in the forthcoming Second \emph{Swift} Ultraviolet/Optical Telescope GRB Afterglow Catalog \citep{Roming2013}.  We detect the at least 119 unique flaring episodes, many of which are previously unreported.  The new method uses the Bayesian Information Criterion to determine the optimal number of breakpoints are needed to provide the best statistical fit to the data.  This method is generic in its application and can be used to find any sort of `feature' which causes the light curve to deviate from the expected decay slope.  We plan to apply this same detection algorithm to the matching set of GRB light curves as observed by the \emph{Swift} X-ray Telescope and publish a catalog of X-ray flares.  Using these two catalogs together we will correlate the X-ray and UV/optical flaring in an attempt to better understand the nature of GRB flaring and ultimately their origin.


\begin{thebibliography}{35}
\expandafter\ifx\csname natexlab\endcsname\relax\def\natexlab#1{#1}\fi

\bibitem[{Barthelmy {et~al.}(2005)Barthelmy, Barbier, Cummings, Fenimore,
  Gehrels, Hullinger, Krimm, Markwardt, Palmer, Parsons, Sato, Suzuki,
  Takahashi, Tashiro, \& Tueller}]{Barthelmy2005}
Barthelmy, S.~D., Barbier, L.~M., Cummings, J.~R., {et~al.} 2005, Space Science
  Reviews, 120, 143

\bibitem[{Burrows {et~al.}(2005{\natexlab{a}})Burrows, Romano, Falcone,
  Kobayashi, Zhang, Moretti, O'Brien, Goad, Campana, Page, Angelini, Barthelmy,
  Beardmore, Capalbi, Chincarini, Cummings, Cusumano, Fox, Giommi, Hill,
  Kennea, Krimm, Mangano, Marshall, M\'{e}sz\'{a}ros, Morris, Nousek, Osborne,
  Pagani, Perri, Tagliaferri, Wells, Woosley, \& Gehrels}]{Burrows2005a}
Burrows, D.~N., Romano, P., Falcone, A., {et~al.} 2005{\natexlab{a}}, Science,
  309, 1833

\bibitem[{Burrows {et~al.}(2005{\natexlab{b}})Burrows, Hill, Nousek, Kennea,
  Wells, Osborne, Abbey, Beardmore, Mukerjee, Short, Chincarini, Campana,
  Citterio, Moretti, Pagani, Tagliaferri, Giommi, Capalbi, Tamburelli,
  Angelini, Cusumano, Br\"{a}uninger, Burkert, \& Hartner}]{Burrows2005}
Burrows, D.~N., Hill, J.~E., Nousek, J.~a., {et~al.} 2005{\natexlab{b}}, Space
  Science Reviews, 120, 165

\bibitem[{Campana {et~al.}(2006)Campana, Tagliaferri, Lazzati, Chincarini,
  Covino, Page, Romano, Moretti, Cusumano, Mangano, Mineo, {La Parola}, Giommi,
  Perri, Capalbi, Zhang, Barthelmy, Cummings, Sakamoto, Burrows, Kennea,
  Nousek, Osborne, O'Brien, Godet, \& Gehrels}]{Campana2006}
Campana, S., Tagliaferri, G., Lazzati, D., {et~al.} 2006, Astronomy and
  Astrophysics, 454, 113

\bibitem[{Chincarini {et~al.}(2009)Chincarini, Margutti, Mao, Pasotti,
  Guidorzi, Covino, \& D‚ÄôAvanzo}]{Chincarini2009}
Chincarini, G., Margutti, R., Mao, J., {et~al.} 2009, Advances in Space
  Research, 43, 1457

\bibitem[{Chincarini {et~al.}(2007)Chincarini, Moretti, Romano, Falcone,
  Morris, Racusin, Campana, Covino, Guidorzi, Tagliaferri, Burrows, Pagani,
  Stroh, Grupe, Capalbi, Cusumano, Gehrels, Giommi, {La Parola}, Mangano,
  Mineo, Nousek, O'Brien, Page, Perri, Troja, Willingale, \&
  Zhang}]{Chincarini2007}
Chincarini, G., Moretti, A., Romano, P., {et~al.} 2007, The Astrophysical
  Journal, 671, 1903

\bibitem[{Chincarini {et~al.}(2010)Chincarini, Mao, Margutti, Bernardini,
  Guidorzi, Pasotti, Giannios, Valle, Moretti, Romano, D'Avanzo, Cusumano, \&
  Giommi}]{Chincarini2010}
Chincarini, G., Mao, J., Margutti, R., {et~al.} 2010, Monthly Notices of the
  Royal Astronomical Society, 406, 2113

\bibitem[{Cusumano {et~al.}(2007)Cusumano, Mangano, Chincarini, Panaitescu,
  Burrows, {La Parola}, Sakamoto, Campana, Mineo, Tagliaferri, Angelini,
  Barthelmy, Beardmore, Boyd, Cominsky, Gronwall, Fenimore, Gehrels, Giommi,
  Goad, Hurley, Immler, Kennea, Mason, Marshall, M\'{e}sz\'{a}ros, Nousek,
  Osborne, Palmer, Roming, Wells, White, \& Zhang}]{Cusumano2007}
Cusumano, G., Mangano, V., Chincarini, G., {et~al.} 2007, Astronomy and
  Astrophysics, 462, 73

\bibitem[{Evans {et~al.}(2009)Evans, Beardmore, Page, Osborne, O'Brien,
  Willingale, Starling, Burrows, Godet, Vetere, Racusin, Goad, Wiersema,
  Angelini, Capalbi, Chincarini, Gehrels, Kennea, Margutti, Morris, Mountford,
  Pagani, Perri, Romano, \& Tanvir}]{Evans2009}
Evans, P.~A., Beardmore, A.~P., Page, K.~L., {et~al.} 2009, Monthly Notices of
  the Royal Astronomical Society, 397, 1177

\bibitem[{Falcone {et~al.}(2006)Falcone, Burrows, Lazzati, Campana, Kobayashi,
  Zhang, Meszaros, Page, Kennea, Romano, Pagani, Angelini, Beardmore, Capalbi,
  Chincarini, Cusumano, Giommi, Goad, Godet, Grupe, Hill, {La Parola}, Mangano,
  Moretti, Nousek, O'Brien, Osborne, Perri, Tagliaferri, Wells, \&
  Gehrels}]{Falcone2006}
Falcone, A.~D., Burrows, D.~N., Lazzati, D., {et~al.} 2006, The Astrophysical
  Journal, 641, 1010

\bibitem[{Falcone {et~al.}(2007)Falcone, Morris, Racusin, Chincarini, Moretti,
  Romano, Burrows, Pagani, Stroh, Grupe, Campana, Covino, Tagliaferri,
  Willingale, \& Gehrels}]{Falcone2007}
Falcone, A.~D., Morris, D., Racusin, J.~L., {et~al.} 2007, The Astrophysical
  Journal, 671, 1921

\bibitem[{Gehrels {et~al.}(2004)Gehrels, Chincarini, Giommi, Mason, Nousek,
  Wells, White, Barthelmy, Burrows, Cominsky, Hurley, Marshall, Meszaros,
  Roming, Angelini, Barbier, Belloni, Campana, Caraveo, Chester, Citterio,
  Cline, Cropper, Cummings, Dean, Feigelson, Fenimore, Frail, Fruchter,
  Garmire, Gendreau, Ghisellini, Greiner, Hill, Hunsberger, Krimm, Kulkarni,
  Kumar, Lebrun, Lloyd‚ÄêRonning, Markwardt, Mattson, Mushotzky, Norris,
  Osborne, Paczynski, Palmer, Park, Parsons, Paul, Rees, Reynolds, Rhoads,
  Sasseen, Schaefer, Short, Smale, Smith, Stella, Tagliaferri, Takahashi,
  Tashiro, Townsley, Tueller, Turner, Vietri, Voges, Ward, Willingale, Zerbi,
  \& Zhang}]{Gehrels2004}
Gehrels, N., Chincarini, G., Giommi, P., {et~al.} 2004, The Astrophysical
  Journal, 611, 1005

\bibitem[{Hill {et~al.}(2006)Hill, Morris, Sakamoto, Sato, Burrows, Angelini,
  Pagani, Moretti, Abbey, Barthelmy, Beardmore, Biryukov, Campana, Capalbi,
  Cusumano, Giommi, Ibrahimov, Kennea, Kobayashi, Ioka, Markwardt, Meszaros,
  O'Brien, Osborne, Pozanenko, Perri, Rumyantsev, Schady, Sharapov,
  Tagliaferri, Zhang, Chincarini, Gehrels, Wells, \& Nousek}]{Hill2006}
Hill, J.~E., Morris, D.~C., Sakamoto, T., {et~al.} 2006, The Astrophysical
  Journal, 639, 303

\bibitem[{Kass \& Raftery(1995)}]{Kass1995}
Kass, R., \& Raftery, A. 1995, Journal of the American Statistical Association,
  90, 773

\bibitem[{Margutti {et~al.}(2010)Margutti, Guidorzi, Chincarini, Bernardini,
  Genet, Mao, \& Pasotti}]{Margutti2010}
Margutti, R., Guidorzi, C., Chincarini, G., {et~al.} 2010, Monthly Notices of
  the Royal Astronomical Society, 406, 2149

\bibitem[{Morgan {et~al.}(2008)Morgan, {Vanden Berk}, Roming, Nousek, Koch,
  Breeveld, de~Pasquale, Holland, Kuin, Page, \& Still}]{Morgan2008}
Morgan, A.~N., {Vanden Berk}, D.~E., Roming, P. W.~A., {et~al.} 2008, The
  Astrophysical Journal, 683, 913

\bibitem[{Morris(2008)}]{Morris2008}
Morris, D.~C. 2008, PhD thesis, The Pennsylvania State University

\bibitem[{Morris {et~al.}(2007)Morris, Reeves, Pal‚Äôshin, Garczarczyk,
  Falcone, Burrows, Krimm, Galante, Gaug, Golenetskii, Mizobuchi, Pagani,
  Stamerra, Teshima, Beardmore, Godet, \& Gehrels}]{Morris2007}
Morris, D.~C., Reeves, J., Pal‚Äôshin, V., {et~al.} 2007, The Astrophysical
  Journal, 654, 413

\bibitem[{Nousek {et~al.}(2006)Nousek, Kouveliotou, Grupe, Page, Granot,
  Ramirez‚ÄêRuiz, Patel, Burrows, Mangano, Barthelmy, Beardmore, Campana,
  Capalbi, Chincarini, Cusumano, Falcone, Gehrels, Giommi, Goad, Godet,
  Hurkett, Kennea, Moretti, O'Brien, Osborne, Romano, Tagliaferri, \&
  Wells}]{Nousek2006}
Nousek, J.~A., Kouveliotou, C., Grupe, D., {et~al.} 2006, The Astrophysical
  Journal, 642, 389

\bibitem[{Oates {et~al.}(2009)Oates, Page, Schady, de~Pasquale, Koch, Breeveld,
  Brown, Chester, Holland, Hoversten, Kuin, Marshall, Roming, Still, {Vanden
  Berk}, Zane, \& Nousek}]{Oates2009}
Oates, S.~R., Page, M.~J., Schady, P., {et~al.} 2009, Monthly Notices of the
  Royal Astronomical Society, 395, 490

\bibitem[{O'Brien {et~al.}(2006)O'Brien, Willingale, Osborne, Goad, Page,
  Vaughan, Rol, Beardmore, Godet, Hurkett, Wells, Zhang, Kobayashi, Burrows,
  Nousek, Kennea, Falcone, Grupe, Gehrels, Barthelmy, Cannizzo, Cummings, Hill,
  Krimm, Chincarini, Tagliaferri, Campana, Moretti, Giommi, Perri, Mangano, \&
  LaParola}]{OBrien2006}
O'Brien, P.~T., Willingale, R., Osborne, J.~P., {et~al.} 2006, The
  Astrophysical Journal, 647, 1213

\bibitem[{Piro {et~al.}(1998)Piro, Amati, Antonelli, Butler, Costa, Cusumano,
  Feroci, Frontera, Heise, Zand, \& Others}]{Piro1998}
Piro, L., Amati, L., Antonelli, L., {et~al.} 1998, Astronomy \& Astrohpysics,
  331, 41

\bibitem[{Piro {et~al.}(2005)Piro, {De Pasquale}, Soffitta, Lazzati, Amati,
  Costa, Feroci, Frontera, Guidorzi, {in ‚Äôt Zand}, Montanari, \&
  Nicastro}]{Piro2005}
Piro, L., {De Pasquale}, M., Soffitta, P., {et~al.} 2005, The Astrophysical
  Journal, 623, 314

\bibitem[{Romano {et~al.}(2006)Romano, Moretti, Banat, Burrows, Campana,
  Chincarini, Covino, \& Malesani}]{Romano2006}
Romano, P., Moretti, A., Banat, P.~L., {et~al.} 2006, Astronomy, 68, 59

\bibitem[{Roming {et~al.}(2013)Roming, Koch, \& Oates}]{Roming2013}
Roming, P. W.~A., Koch, T.~S., \& Oates, S.~R. 2013, The Astrophysical Journal

\bibitem[{Roming {et~al.}(2006{\natexlab{a}})Roming, {Vanden Berk}, Hunsberger,
  Page, Mason, Marshall, \& Boyd}]{Roming2006a}
Roming, P. W.~A., {Vanden Berk}, D.~E., Hunsberger, S.~D., {et~al.}
  2006{\natexlab{a}}, Il Nuovo Cimento B, 121, 1239

\bibitem[{Roming {et~al.}(2000)Roming, Townsley, Nousek, Altimore, Case,
  Hunsberger, Koch, Mason, Carter, Cropper, Hancock, Huckle, Kennedy,
  Mclelland, Smith, Killough, Ho, Mary, \& Rh}]{Roming2000}
Roming, P. W.~A., Townsley, L.~K., Nousek, J.~A., {et~al.} 2000, Proc. SPIE,
  4140, 76

\bibitem[{Roming {et~al.}(2004)Roming, Hunsberger, Mason, Nousek, Broos,
  Carter, Hancock, Huckle, Kennedy, Killough, Koch, McLelland, Pryzby, Smith,
  Soto, Stock, Boyd, \& Still}]{Roming2004}
Roming, P. W.~A., Hunsberger, S.~D., Mason, K.~O., {et~al.} 2004, Proc. SPIE,
  5165, 262

\bibitem[{Roming {et~al.}(2005)Roming, Kennedy, Mason, Nousek, Ahr, Bingham,
  Broos, Carter, Hancock, Huckle, Hunsberger, Kawakami, Killough, Koch,
  Mclelland, Smith, Smith, Soto, Boyd, Breeveld, Holland, Ivanushkina, Pryzby,
  Still, \& Stock}]{Roming2005}
Roming, P. W.~A., Kennedy, T.~E., Mason, K.~O., {et~al.} 2005, Space Science
  Reviews, 120, 95

\bibitem[{Roming {et~al.}(2006{\natexlab{b}})Roming, {Vanden Berk}, Pal‚Äôshin,
  Pagani, Norris, Kumar, Krimm, Holland, Gronwall, Blustin, Zhang, Schady,
  Sakamoto, Osborne, Nousek, Marshall, Meszaros, Golenetskii, Gehrels,
  Frederiks, Campana, Burrows, Boyd, Barthelmy, \& Aptekar}]{Roming2006b}
Roming, P. W.~A., {Vanden Berk}, D., Pal‚Äôshin, V., {et~al.}
  2006{\natexlab{b}}, The Astrophysical Journal, 651, 985

\bibitem[{Roming {et~al.}(2009)Roming, Koch, Oates, Porterfield, {Vanden Berk},
  Boyd, Holland, Hoversten, Immler, Marshall, Page, Racusin, Schneider,
  Breeveld, Brown, Chester, Cucchiara, {De Pasquale}, Gronwall, Hunsberger,
  Kuin, Landsman, Schady, \& Still}]{Roming2009}
Roming, P. W.~A., Koch, T.~S., Oates, S.~R., {et~al.} 2009, The Astrophysical
  Journal, 690, 163

\bibitem[{Schwarz(1978)}]{Schwarz1978}
Schwarz, G. 1978, The Annals of Statistics, 6, 461

\bibitem[{Swenson {et~al.}(2010)Swenson, Maxham, Roming, Schady, Vetere, Zhang,
  Zhang, Holland, Kennea, Kuin, Oates, Page, \& {De Pasquale}}]{Swenson2010}
Swenson, C.~A., Maxham, A., Roming, P. W.~A., {et~al.} 2010, The Astrophysical
  Journal, 718, L14

\bibitem[{Zeileis {et~al.}(2003)Zeileis, Kleiber, Kramer, \&
  Hornik}]{Zeileis2003}
Zeileis, A., Kleiber, C., Kramer, W., \& Hornik, K. 2003, Computational
  Statistics \& Data Analysis, 44, 109

\bibitem[{Zeileis {et~al.}(2002)Zeileis, Leisch, Hornik, \&
  Kleiber}]{Zeileis2002}
Zeileis, A., Leisch, F., Hornik, K., \& Kleiber, C. 2002, Journal of
  Statistical Software, 7, 1

\end{thebibliography}

\appendix
\section{Demonstration of \emph{breakpoints} function on simulated data}

To verify the effectiveness of the breakpoints function, we performed a series of tests on simulated UVOT data.  Starting with a simple power-law light curve, we induced a number of flares on the light curve and attempted to detect those flares using the method previously described.  In order to test the ability of the breakpoints function to detect a wide variety of flares we varied the $T_{start}$ time, the amplitude, the duration of the induced flares, as well as the slope of the underlying light curve.  All of our analysis is done on the residuals of the best fit to the light curve.

We use the criteria of $BIC_{i} - BIC_{min} > 6$, or `Strong' evidence from \citet{Kass1995}, to determine the appropriate number of breakpoints to assign to a specific light curve.  We then attempt to group breakpoints together to form individual flares based on the relative position of the potential breakpoints relative to each other and relative to the underlying light curve.  Our measure of confidence is determined by the number of times a specific breakpoint was identified in the course of the 10,000 Monte Carlo simulations, and also satisfied the criteria of $BIC_{i} - BIC_{min} > 6$.

\begin{figure}
	\includegraphics[width=\textwidth]{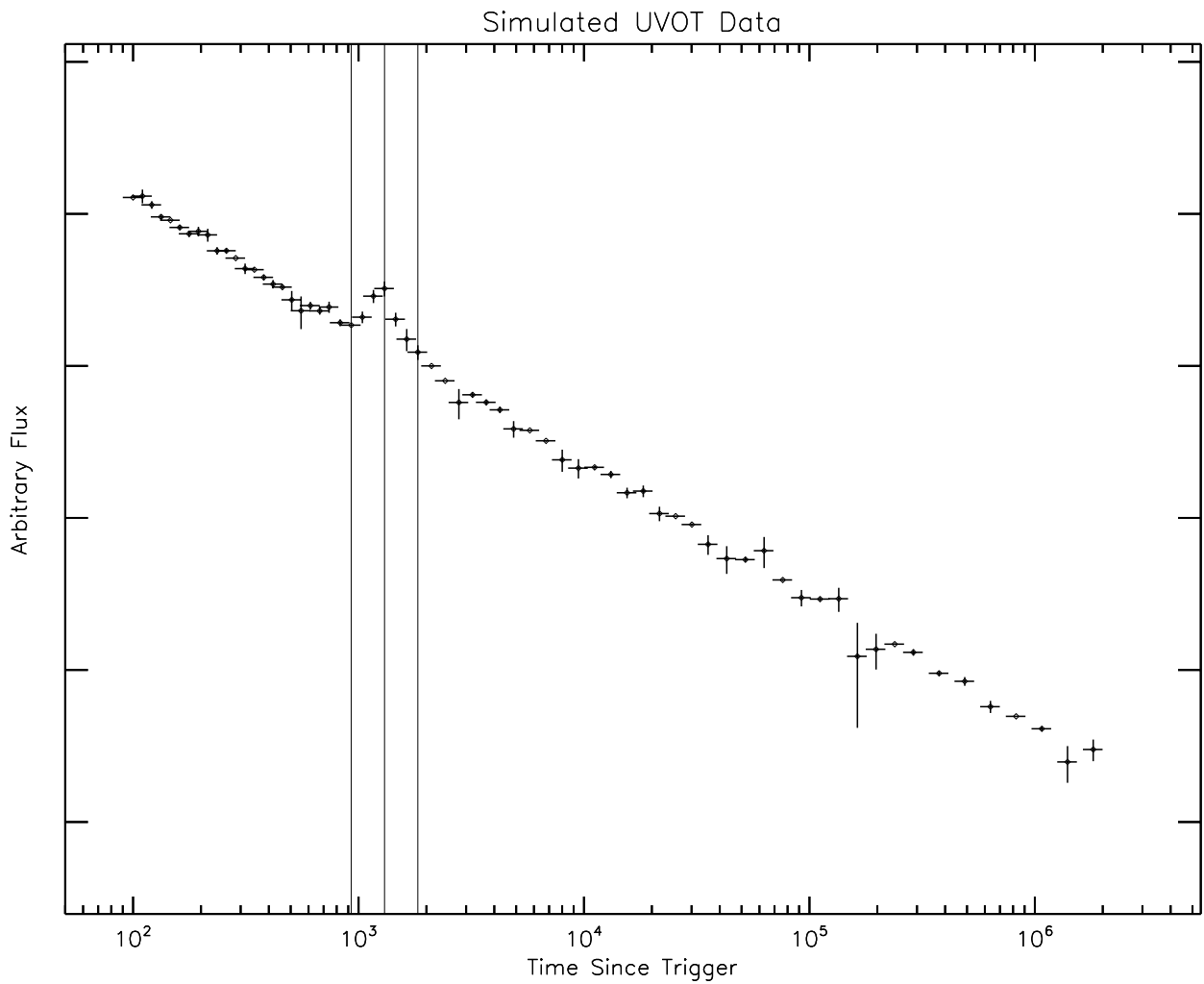}
	\caption{Simulated UVOT data showing an optimal flare.  All three components of the flare, $T_{start}$, $T_{peak}$ and $T_{stop}$, are detected and identified.}
	\label{fig:SimulatedDataOptimal}
\end{figure}

As previously mentioned, the ideal case is when the amplitude and duration of the flare are both large enough to make $T_{start}$, $T_{peak}$, and $T_{stop}$ easily identifiable.  Figure~\ref{fig:SimulatedDataOptimal} shows an example of simulated UVOT data with an induced flare that meets this criteria, and shows the potential breakpoints found by the code as vertical lines on top of the data.  Analysis of the potential breakpoints in relation to each other and to the underlying light curve show that they collectively form a single flare with the three potential breakpoints marking the approximate times of $T_{start}$, $T_{peak}$, and $T_{stop}$.  As with all the identified flares, the values assigned to $T_{start}$ and $T_{stop}$ are lower and upper limits of the actual values, with the last data point before the flare that is still well fit by the underlying light curve assigned to $T_{start}$ and the first data point after the flare well fit by the underlying light curve assigned to $T_{stop}$.  In the case of a well sampled light curve and flare, like that shown in Figure~\ref{fig:SimulatedDataOptimal}, the limits given by $T_{start}$ and $T_{stop}$  do not differ significantly from calculated values derived from fitting the flare.  The difference is less by than the size of the exposure bin for data points we identify as $T_{start}$ and $T_{stop}$.  We therefore see no need to differentiate our methodology and to fit those few flares that can be fit using a function.  For this flare, the measure of confidence for each individual breakpoint is very high.  Specifically, the breakpoint associated with $T_{peak}$ was identified in all 10,000 iterations.  In this case we are therefore 100\% confident in the presence of a flare despite not detecting all three components at 100\% confidence.  We assign the overall confidence of the flare to be 10,000, reflecting this certainty.  This example is truly the exceptional case, as UV/optical flares are rarely observed with such strength.

\begin{figure}
	\includegraphics[width=\textwidth]{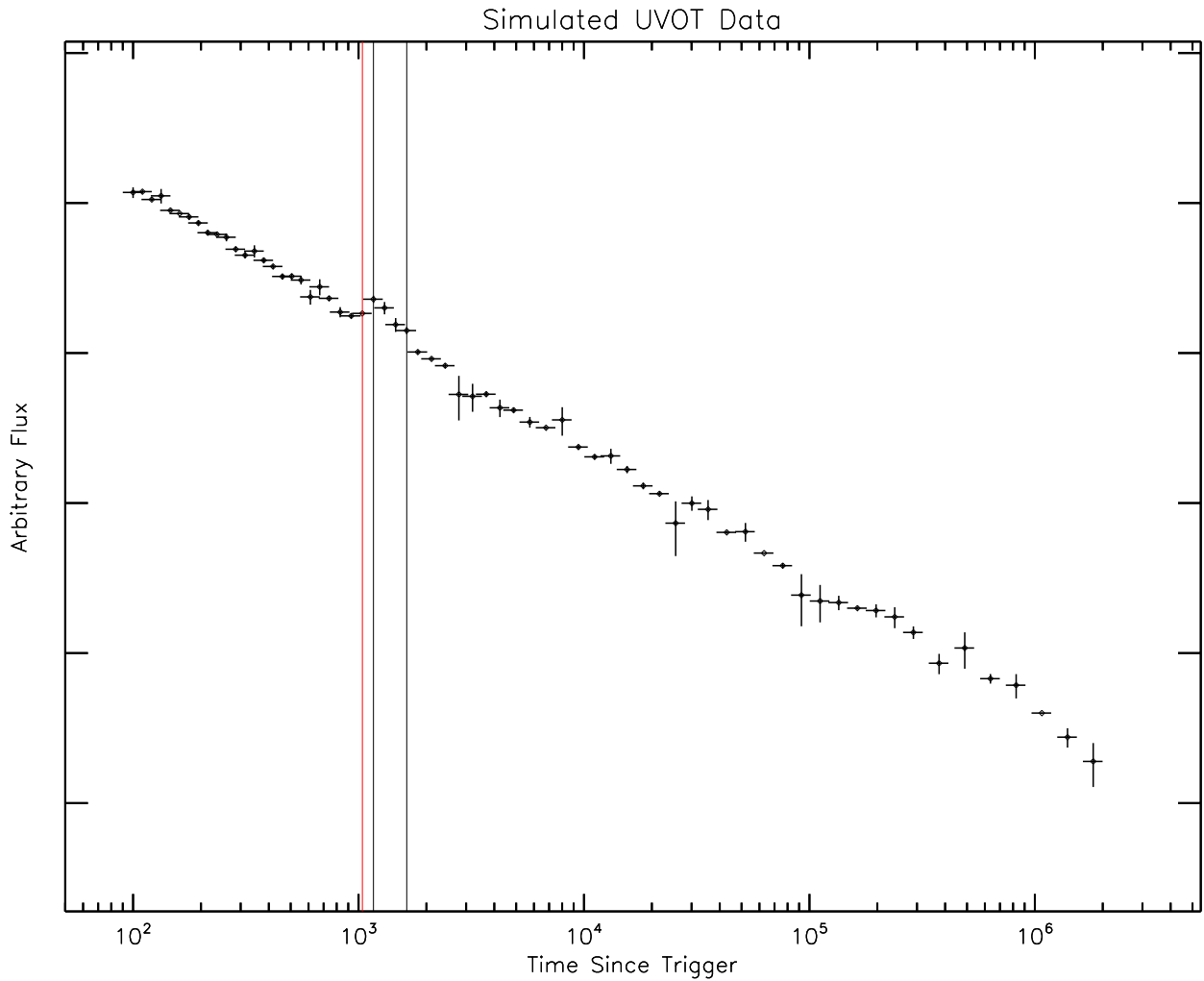}
	\caption{Simulated UVOT data showing a flare with a small amplitude and duration.  Due to the abrupt rise to the peak of the flare, the code does not identify a unique point as being associated with the start of the flare.  We assign the first point prior to $T_{peak}$ to be $T_{start}$.  The data point assigned as $T_{start}$ is identified by the red line.}
	\label{fig:SimulatedDataNonOptimal}
\end{figure}

A more typical size flare is shown in Figure~\ref{fig:SimulatedDataNonOptimal}.  In this case, the vertical lines again show the positions of potential breakpoints found by the flare finding code.  Further analysis shows that these two breakpoints again form a single flare, however, only the approximate $T_{peak}$ and $T_{stop}$ times have been identified.  A potential $T_{start}$ breakpoint was never identified during the 10,000 Monte Carlo simulations.  This does not mean that the code failed to properly identify the flare, but rather shows that even at low significance the code is able to detect flares, however all the individual components of the flare may not be detected.  In the case of the flare in Figure~\ref{fig:SimulatedDataNonOptimal}, the rise to the observed peak of the flare is short enough, due to a combination of a shorter flare duration and a small peak amplitude, that there are no observed data points during the rise of the flare.  The first data point found by the flare finding code that deviates from the underlying light curve is the observed peak of the flare.  In cases such as these the code picks the nearest observed data point before the observed flare peak as being the lower limit for the $T_{start}$.  In this case, the assigned time for $T_{start}$ is once again very close to the actual start of the flare.  This flare is given a confidence measure of 7,434, which is the confidence measure associated with the observed flare peak.

\begin{figure}
	\includegraphics[width=\textwidth]{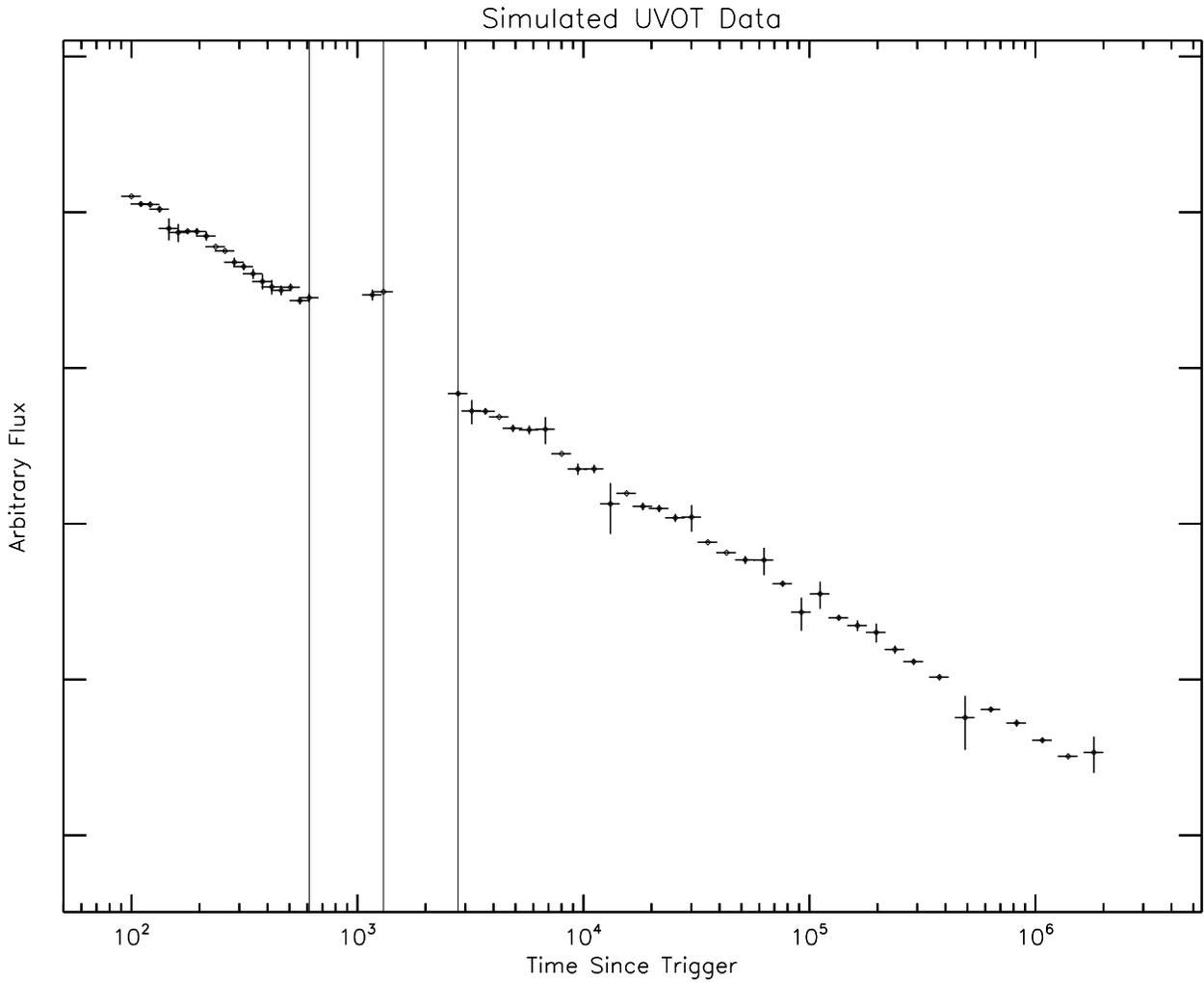}
	\caption{Simulated UVOT data showing the influence of observing gaps on the flare finding code.  An elevated flux level is identified as a potential $T_{peak}$, however the beginning and ending of the flare are not observed.  The first data point prior to the observing gap is designated as $T_{start}$ and the first data point after the second observing gap is designated as $T_{stop}$.}
	\label{fig:SimulatedDataGap}
\end{figure}

In both of the previous examples, the sample of the light curves was continuous and uninterrupted.  However, this is never the case with actual data.  Our data from the Second \emph{UVOT} GRB Catalog \citep{Roming2013} are influenced, at the very least, by the fact that the \emph{Swift} satellite has a 96 minute orbit and that any target on the sky will be unobservable for $>$50\% of the orbit.  Additionally any number of other factors including the observing of higher priority targets and conflicts with spacecraft observing constraints, have produced light curves with uneven sampling and occasional large gaps.  These gaps are the reason that we have decided to provide limits on the values of $T_{start}$ and $T_{stop}$ and to use the observed time of peak flux when reporting $T_{peak}$.  Figure~\ref{fig:SimulatedDataGap} shows the results from running the flare finding code on the same basic light curve and induced flares as in Figure~\ref{fig:SimulatedDataOptimal}, but with an observing gap overlapping the beginning of the flare.  The code once again identifies three distinct data points that are potential breakpoints.  Analysis of the data points shows that we are unable to determine whether they are all associated with a single flaring event due to gaps in the light curve.  Not only are we uncertain whether the first and last potential breakpoints correspond to a $T_{start}$ and $T_{stop}$ of a given flare or multiple flares, we are also no longer confident that the point identified as the peak observed flux is an accurate approximation of the actual $T_{peak}$ or the peak flux level reached during the flare(s).  Again we assign the last data point before the flaring period that was well fit by the underlying light curve as $T_{start}$ (in this case the last observed data point before the observing gap), and do the same for $T_{stop}$ by assigning the first data point after the flaring period well fit by the underlying light curve (in this case the first data point after the second observing gap).  Because we did not observe the majority of the flare and are uncertain in the actual peak flux achieved, the flux ratio we report becomes a lower limit, based on the peak observed flux and the flux of the underlying light curve at the time of the observed peak.  This flare is given a confidence measure of 9,784, which is again the confidence measure associated with the observed peak.

In each of these examples we have only shown cases where there are no additional potential breakpoints other than those associated with either $T_{start}$, $T_{peak}$ or $T_{stop}$.  Occasionally the code does find more than three breakpoints for a single flare, specifically in situations where there is poor data sampling during the peak of the flare.  In these cases, if no single data point stands out as a peak, the code will identify points to either side of the peak as being potential peak candidates, resulting in four breakpoints for a single flare.  In these cases we assign the data point with the larger flux to be the peak flux time used in our calculations of $\Delta t/t$ and the flux ratio.  It should be noted, that this same series of four breakpoints could be the result of two individual flares occurring in quick succession.  If the timing resolution is larger than the $\Delta t/t$ of the flares, the code will not be able to correctly identify two individual flares, but will rather identify a single broad flaring period.

\begin{deluxetable}{lrrrrrr}
\tabletypesize{\scriptsize}
\tablecaption{Optical GRB flares from the Second UVOT GRB Catalog}
\tablewidth{0pt}
\tablecolumns{7}
\tablehead{
\colhead{Source Name} &
\colhead{$T_{peak}$*} &
\colhead{$T_{start}$ lower limit*} &
\colhead{$T_{stop}$ upper limit*} &
\colhead{$\Delta$t/t} &
\colhead{Flux Ratio} &
\colhead{Confidence} \\
\colhead{} &
\colhead{(s)} &
\colhead{(s)} &
\colhead{(s)} &
\colhead{} &
\colhead{lower limit} &
\colhead{}
}
\startdata
GRB050319 & 333.58 & 272.87 & 428.21 & 0.47  & 0.83  & 0.5075 \\
GRB050319 & 1061.79 & 927.06 & 1208.41 & 0.26  & 1.19  & 0.3259 \\
GRB050319 & 802.84 & 739.17 & 927.06 & 0.23  & 0.31  & 0.252 \\
GRB050505 & 3181.79 & 3181.79 & 8065.87 & 1.54  & 2.11  & 0.9408 \\
GRB050525 & 271.40 & 257.75 & 299.17 & 0.15  & 0.56  & 0.9644 \\
GRB050525 & 608.92 & 510.85 & 636.67 & 0.21  & 1.24  & 0.4149 \\
GRB050525 & 186.92 & 172.68 & 214.72 & 0.22  & 1.03  & 0.9651 \\
GRB050712 & 787496.25 & 714478.38 & 844068.38 & 0.16  & 26.48 & 0.9601 \\
GRB050721 & 494.74 & 423.97 & 508.92 & 0.17  & 0.54  & 0.4482 \\
GRB050801 & 398.35 & 383.12 & 439.89 & 0.14  & 1.24  & 0.885 \\
GRB050801 & 341.22 & 327.61 & 355.35 & 0.08  & 0.72  & 0.8037 \\
GRB050802 & 887.98 & 803.49 & 1104.62 & 0.34  & 0.91  & 0.5038 \\
GRB050802 & 1428.47 & 1299.30 & 1492.69 & 0.14  & 0.70  & 0.4371 \\
GRB050815 & 131.78 & 99.32 & 145.33 & 0.35  & 1.64  & 0.5691 \\
GRB050824 & 94236.29 & 81838.89 & 203483.48 & 1.29  & 11.65 & 0.7743 \\
GRB050908 & 368.58 & 214.89 & 435.09 & 0.60  & 1.59  & 0.9147 \\
GRB051117A & 610.93 & 555.99 & 773.26 & 0.36  & 0.85  & 0.616 \\
GRB060206 & 129.90 & 93.51 & 1444.54 & 10.40 & 1.67  & 0.6495 \\
GRB060313 & 524.12 & 470.21 & 795.27 & 0.62  & 1.39  & 0.7843 \\
GRB060428B & 145182.38 & 141352.78 & 174712.64 & 0.23  & 8.90  & 0.568 \\
GRB060512 & 448.07 & 432.01 & 4019.50 & 8.01  & 2.25  & 0.817 \\
GRB060526 & 262.77 & 242.76 & 272.78 & 0.11  & 1.08  & 0.9724 \\
GRB060526 & 192.71 & 177.44 & 242.76 & 0.34  & 0.75  & 0.9178 \\
GRB060526 & 292.79 & 282.79 & 312.81 & 0.10  & 0.04  & 0.5578 \\
GRB060604 & 203.71 & 193.70 & 213.57 & 0.10  & 0.44  & 0.4402 \\
GRB060708 & 108.65 & 98.65 & 128.67 & 0.28  & 0.34  & 0.6744 \\
GRB060729 & 18581.79 & 12568.86 & 31109.46 & 1.00  & 0.37  & 0.9989 \\
GRB060904B & 256.86 & 234.94 & 284.31 & 0.19  & 2.13  & 0.8475 \\
GRB060912 & 225.13 & 209.74 & 245.14 & 0.16  & 0.41  & 0.7158 \\
GRB060912 & 375.24 & 365.23 & 395.25 & 0.08  & 0.70  & 0.5805 \\
GRB060912 & 285.17 & 265.16 & 305.19 & 0.14  & 0.24  & 0.5041 \\
GRB061021 & 222.52 & 192.50 & 232.53 & 0.18  & 0.54  & 0.7412 \\
GRB061021 & 4274.53 & 452.69 & 4683.61 & 0.99  & 0.31  & 0.5813 \\
GRB061021 & 5295.80 & 4683.61 & 5298.24 & 0.12  & 0.27  & 0.5805 \\
GRB070208 & 433659.09 & 427659.38 & 480022.66 & 0.12  & 1.52  & 0.673 \\
GRB070318 & 246.45 & 226.44 & 256.46 & 0.12  & 0.16  & 0.2538 \\
GRB070318 & 191458.19 & 142816.11 & 296319.50 & 0.80  & 2.63  & 0.3125 \\
GRB070420 & 1939.71 & 1750.87 & 5792.08 & 2.08  & 2.03  & 0.6216 \\
GRB070518 & 273.72 & 243.70 & 317.92 & 0.27  & 1.63  & 0.8283 \\
GRB070518 & 193.65 & 177.69 & 206.81 & 0.15  & 1.24  & 0.5656 \\
GRB070518 & 117.81 & 107.80 & 137.83 & 0.25  & 0.24  & 0.4614 \\
GRB070611 & 4733.36 & 3347.02 & 10492.06 & 1.51  & 0.78  & 0.9438 \\
GRB070616 & 1011.36 & 846.23 & 1149.26 & 0.30  & 0.90  & 0.9267 \\
GRB070616 & 787.52 & 468.17 & 816.63 & 0.44  & 0.47  & 0.8939 \\
GRB070721B & 275.26 & 255.24 & 285.27 & 0.11  & 1.17  & 0.9697 \\
GRB071031 & 1118.38 & 1102.01 & 1158.87 & 0.05  & 1.18  & 0.7647 \\
GRB071031 & 576.25 & 546.56 & 842.05 & 0.51  & 1.32  & 0.74 \\
GRB071031 & 11666.85 & 7596.43 & 14239.37 & 0.57  & 1.60  & 0.6842 \\
GRB071112C & 595.43 & 572.52 & 636.93 & 0.11  & 1.57  & 0.7843 \\
GRB071112C & 245.42 & 215.39 & 285.45 & 0.29  & 1.38  & 0.706 \\
GRB071112C & 18094.00 & 12735.36 & 45965.98 & 1.84  & 55.43 & 0.6809 \\
GRB080212 & 156.79 & 126.10 & 178.16 & 0.33  & 1.82  & 0.7306 \\
GRB080212 & 266.88 & 223.14 & 295.85 & 0.27  & 0.90  & 0.535 \\
GRB080212 & 356.95 & 340.57 & 378.72 & 0.11  & 1.09  & 0.2891 \\
GRB080303 & 573.73 & 512.75 & 620.90 & 0.19  & 2.07  & 0.7797 \\
GRB080303 & 223.47 & 193.45 & 256.43 & 0.28  & 0.76  & 0.5921 \\
GRB080319B & 252.59 & 232.58 & 262.60 & 0.12  & 0.10  & 0.6566 \\
GRB080319C & 1289.59 & 1194.14 & 1394.45 & 0.16  & 1.92  & 0.8361 \\
GRB080330 & 138.31 & 128.30 & 148.32 & 0.14  & 0.46  & 0.7335 \\
GRB080413B & 428.29 & 412.44 & 453.43 & 0.10  & 1.18  & 0.712 \\
GRB080520 & 192.45 & 170.56 & 332.62 & 0.84  & 0.65  & 0.4985 \\
GRB080703 & 146.84 & 136.83 & 166.86 & 0.20  & 0.26  & 0.8849 \\
GRB080721 & 123.54 & 123.54 & 133.55 & 0.08  & 0.08  & 0.6278 \\
GRB080721 & 300.46 & 290.45 & 330.48 & 0.13  & 0.26  & 0.6111 \\
GRB080804 & 482.51 & 412.46 & 532.55 & 0.25  & 0.46  & 0.8254 \\
GRB080810 & 113.06 & 103.06 & 133.09 & 0.27  & 0.16  & 0.9201 \\
GRB080810 & 229.12 & 199.09 & 289.17 & 0.39  & 0.17  & 0.7133 \\
GRB080905B & 286.85 & 276.84 & 306.56 & 0.10  & 1.01  & 0.5412 \\
GRB080905B & 507.00 & 476.98 & 527.02 & 0.10  & 0.36  & 0.3713 \\
GRB080906 & 256.26 & 241.19 & 284.02 & 0.17  & 4.86  & 0.6933 \\
GRB080913 & 12538.61 & 6082.72 & 14589.25 & 0.68  & 2.49  & 0.7557 \\
GRB080913 & 551292.31 & 512841.41 & 887145.81 & 0.68  & 2.34  & 0.5695 \\
GRB080916A & 470.21 & 460.21 & 490.23 & 0.06  & 1.26  & 0.7993 \\
GRB080916A & 370.14 & 360.14 & 390.16 & 0.08  & 1.47  & 0.5295 \\
GRB080928 & 4329.05 & 2090.70 & 4944.90 & 0.66  & 1.18  & 0.8933 \\
GRB080928 & 247.11 & 217.09 & 257.12 & 0.16  & 1.88  & 0.7826 \\
GRB081007 & 253.00 & 206.98 & 273.03 & 0.26  & 0.89  & 0.6911 \\
GRB081007 & 433.15 & 413.13 & 513.20 & 0.23  & 0.60  & 0.5528 \\
GRB081008 & 262.09 & 243.09 & 302.12 & 0.23  & 0.14  & 0.7585 \\
GRB081008 & 1266.86 & 1227.41 & 1291.19 & 0.05  & 1.33  & 0.7367 \\
GRB081029 & 15712.66 & 8659.63 & 22090.42 & 0.85  & 0.32  & 0.8647 \\
GRB081126 & 153.77 & 143.76 & 173.80 & 0.20  & 0.44  & 0.7555 \\
GRB090123 & 1112.77 & 950.67 & 1122.66 & 0.15  & 0.53  & 0.8807 \\
GRB090123 & 668.08 & 608.41 & 707.52 & 0.15  & 0.55  & 0.8018 \\
GRB090123 & 1408.58 & 1368.14 & 1467.41 & 0.07  & 0.42  & 0.7555 \\
GRB090401B & 1117.76 & 1068.56 & 1187.56 & 0.11  & 6.24  & 0.8217 \\
GRB090510 & 147.51 & 132.47 & 157.52 & 0.17  & 1.29  & 0.7921 \\
GRB090529 & 1204.85 & 957.27 & 1688.56 & 0.61  & 2.97  & 0.9025 \\
GRB090530 & 133.67 & 123.66 & 143.68 & 0.15  & 0.19  & 0.8736 \\
GRB090530 & 173.71 & 153.69 & 183.71 & 0.17  & 0.15  & 0.8189 \\
GRB090530 & 634.84 & 595.07 & 756.78 & 0.25  & 1.72  & 0.689 \\
GRB090618 & 715.48 & 705.59 & 764.51 & 0.08  & 0.54  & 0.8776 \\
GRB090618 & 2089.69 & 2079.80 & 2104.10 & 0.01  & 1.47  & 0.7139 \\
GRB090618 & 2597.47 & 2582.00 & 2621.84 & 0.02  & 0.95  & 0.6723 \\
GRB090618 & 666.85 & 641.01 & 691.17 & 0.08  & 0.68  & 0.6045 \\
GRB090618 & 1916.21 & 1906.32 & 1930.64 & 0.01  & 1.27  & 0.55 \\
GRB090618 & 829.04 & 813.52 & 853.39 & 0.05  & 0.44  & 0.5474 \\
GRB090926A & 47722.57 & 46791.30 & 51749.93 & 0.10  & 2.51  & 0.8781 \\
GRB090926A & 81701.3 & 70136.27 & 144308.98 & 0.97  & 1.12  & 0.9078 \\
GRB090926A & 232412.66 & 196569.48 & 248216.91 & 0.22  & 0.65  & 0.6123 \\
GRB091018 & 341.45 & 331.45 & 371.48 & 0.12  & 0.26  & 0.7297 \\
GRB091029 & 1059.17 & 1049.28 & 1147.60 & 0.09  & 0.82  & 0.6614 \\
GRB091029 & 543.66 & 523.65 & 553.48 & 0.05  & 0.79  & 0.5801 \\
GRB091127 & 69405.64 & 68961.88 & 74007.65 & 0.07  & 1.80  & 0.1644 \\
GRB100425A & 24126.33 & 23640.73 & 38746.47 & 0.63  & 2.82  & 0.9208 \\
GRB100425A & 719.77 & 651.76 & 947.33 & 0.41  & 4.12  & 0.8502 \\
GRB100805A & 151.24 & 141.23 & 171.25 & 0.20  & 0.35  & 0.3225 \\
GRB100805A & 409.43 & 392.57 & 423.80 & 0.08  & 1.06  & 0.3222 \\
GRB100805A & 630.03 & 589.59 & 707.20 & 0.19  & 1.56  & 0.2733 \\
GRB100805A & 39387.52 & 29555.01 & 56840.63 & 0.69  & 1.97  & 0.2561 \\
GRB100814A & 224.40 & 214.39 & 244.42 & 0.13  & 0.35  & 0.3692 \\
GRB100814A & 284.46 & 274.45 & 294.47 & 0.07  & 0.46  & 0.3079 \\
GRB100901A & 416.87 & 396.86 & 436.89 & 0.10  & 1.45  & 0.8464 \\
GRB100901A & 228.79 & 208.77 & 238.80 & 0.13  & 1.55  & 0.2395 \\
GRB100901A & 17679.91 & 10070.17 & 21634.66 & 0.65  & 0.27  & 0.2295 \\
GRB100906A & 445.62 & 424.48 & 465.41 & 0.09  & 3.22  & 0.7149 \\
GRB101017A & 180.65 & 144.90 & 215.47 & 0.39  & 1.78  & 0.7994 \\
GRB101117B & 164.51 & 154.50 & 174.52 & 0.12  & 0.47  & 0.512 \\
GRB101117B & 307.99 & 234.39 & 328.00 & 0.30  & 0.62  & 0.3753 \\
\enddata
\tablecomments{Flares are listed in chronological order by GRB date, then sorted by confidence.  *All times are relative to the time of the initial burst trigger.  $\Delta t/t$ is calculated as ($T_{stop} - T_{start})/T_{peak}$.  $T_{start}$ and $T_{stop}$ are lower and upper limits, respectively.  Flux Ratio is calculated as the flux at the flare peak time divided by the extrapolated flux of the underlying light curve at the same time, normalized using the flux of the underlying light curve, and is a lower limit of the actual peak flux ratio.  The confidence measure represents the fraction of times the flare was identified during the 10,000 Monte Carlo simulations.  }
\label{tab:Flaretable}
\end{deluxetable}

\end{document}